\documentstyle[twocolumn,aps,prl,floats,epsf,multirow]{revtex}

\begin{document}

\def\pt{$ p_{t}$ } \def\xf{$ x_F$ } \def\rt{$ \sqrt\tau$ }
\def\u{$\Upsilon$ } \def\JP{$\psi'$ } \def\J{$J/\psi$ }
\def\up{$\Upsilon^{\prime}$ } \def\upp{$\Upsilon^{\prime\prime}$ }
\def\D{$^2H$ }

\draft

\wideabs{

\title{Energy loss of fast quarks in nuclei}                                
\author{
J.M.~Moss$^a$,
G.T.~Garvey$^a$,
M.B.~Johnson$^a$
M.J.~Leitch$^a$,
P.L.~McGaughey$^a$,
J.C.~Peng$^a$,,
B.Z.~Kopeliovich$^b$,
I.K.~Potashnikova$^b$,
\\ \vspace*{5pt}
}
\address{
$^a$Los Alamos National Laboratory, Los Alamos, NM 87545\\
$^b$Max-Plank-Institut f\"ur Kernphysik, 69029 Heidelberg, Germany\\
}
\date{\today}

\maketitle

\begin{abstract}
We report an analysis of the nuclear dependence of the yield of
Drell-Yan (DY) dimuons from the 800 GeV/c proton bombardment of
$^2H$, C, Ca, Fe, and W targets.  A light-cone formulation of the DY
process is employed in the rest frame of the nucleus. In this frame,
for $x_2\ll x_1$, DY production appears as bremsstrahlung of a virtual
photon followed by decay into dileptons. We treat the two sources of
nuclear suppression, energy loss and shadowing, in a consistent
formulation. Shadowing, involving no free parameters, is calculated
within the light-cone dipole
formalism. Initial-state energy loss, the
only unknown in the problem, is determined from a fit to the
nuclear-dependence ratio versus $x_1$. With the assumption of constant
energy loss per unit path length, we find
$-dE/dz = 2.32 \pm 0.52\pm 0.5$ GeV/fm.
This is the first observation of a nonzero energy loss of partons
traveling in nuclear environment.
                                                                        
\end{abstract}
} 

\begin{figure}[thb]
  \begin{center}
    \mbox{\epsfxsize=2.5in\epsffile{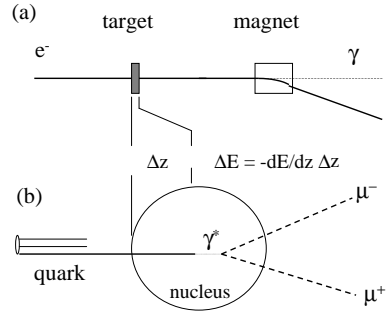}}
    \vspace*{+9pt}                                        
  \end{center}
  \caption{Schematic measurements of (a) energy loss of a fast electron 
beam traversing a thin target, and (b) of a fast quark beam losing 
energy on the front side of a target nucleus. In (b) the energy of the 
initial quark is reconstructed from the four momenta of the lepton pair 
created in the final state.}                           
  \label{eloss}
\end{figure}

\newpage
\noindent{\bf Introduction}
Quarks should lose energy in traversing nuclear matter -- but not very
much. A commonly cited estimate is $-dE/dz\approx\kappa$, where         
$\kappa\approx$ 1 GeV/fm is the QCD string tension. Before discussing quark
energy loss it is instructive to consider the analogous problem in QED,
the energy loss of relativistic electrons passing through solid targets.
Fig.~\ref{eloss}a is an accurate representation of a real experiment to measure 
$dE/dz$. In spite of its conceptual simplicity it was not until 1995\cite{anthony} 
that an accurate measurement of the energy loss in dense matter was made for a 
highly-relativistic electron beam. The experiment confirmed a 
prediction made forty years earlier, now termed the Landau-Pomeranchuk-Migdal 
(LPM) effect\cite{lp,migdal}. In QED the LPM effect is a suppression of 
bremsstrahlung caused by a quantum-mechanical interference between 
different scattering centers. In QCD there is an interesting analogue to 
the LPM effect \cite{bsz} to which much theoretical attention has been devoted in 
recent years. We will return to this point later.

Since, unlike electrons and photons, neither quarks nor gluons travel long 
distances, the QCD {\it gedanken} energy-loss experiment needs an alternative 
realization. A feasible conceptual picture for measurement of quark energy loss is 
given in Fig.~\ref{eloss}b. A quark from an incoming hadron at the left loses 
energy in a nucleus, then undergoes the Drell-Yan (DY) process\cite{drellyan}
producing a lepton pair from the electromagnetic annihilation of the beam quark with 
a target antiquark, $q+\bar q\rightarrow \gamma^*\rightarrow l^+l^-$.
Measurement of the four momenta of the leptons allows reconstruction of 
the momenta of the colliding quark and antiquark. But how much energy 
did the quark have in the first place? That question cannot be answered 
for a particular collision, but the average effect can be deduced by comparing 
DY production from a nucleus to that from a nucleon target where there is 
no energy loss.

\section{Shadowing and energy loss}

The suppression of the cross section for deeply-inelastic lepton 
scattering (DIS) on heavy nuclear targets at small Bjorken-$x$ is known 
as nuclear shadowing\cite{gst}. It is a well-characterized phenomenon, with onset 
for $x\leq 0.07$. In the infinite momentum frame, commonly employed 
in the description of DIS, shadowing can be visualized as the recombination of 
small-$x$ partons whose longitudinal extent exceeds the 
internucleon spacing. Viewed in the target rest frame a different (but 
equivalent) picture of shadowing emerges. Here one focuses on the 
structure of the photon, and its virtual fluctuations into $q\bar q$ 
states which can interact with the target. Small $x$ corresponds to 
fluctuations of the virtual photon whose coherence length exceeds the 
internucleon spacing: they are hence absorbed by more than one nucleon.

Shadowing should also affect hard hadronic processes. The only experimental 
evidence to date is a suppression of the DY cross section on 
heavy targets observed in Fermilab experiments, E772\cite{aldedy} and 
E866\cite{e866}. But is this energy loss or shadowing or some 
combination of the two? Because the two effects can lead to an 
apparently similar nuclear suppression of the DY cross section, it is 
necessary to appeal to a consistently formulated description of both 
effects in order to analyze experimental data. This consistency was 
not required in previous analyses of DY data for quark energy 
loss\cite{e866,gm}.

\section{Drell-Yan process in the target rest frame}

For those used to the usual description of the DY process, 
where a quark and antiquark collide to produce a virtual photon at rest, 
the target rest frame (TRF) view \cite{hir,bhq,johnson} is downright 
strange. In the 
TRF an energetic quark from the incident hadron undergoes continual 
fluctuations into a virtual photon and a residual quark.
The lepton pair results from the decay of the virtual 
photon when the residual quark interacts with the target. This picture, 
which is most useful for DY production at small $x_2$, makes
no explicit reference to antiquarks in the target. 

The TRF analysis begins with DY production from a nucleon target, $^2H$ 
in the case of E772. An incident quark with momentum 
fraction $x_q$ emits a virtual photon that carries a fraction 
$x_1^q =x_1/x_q$ of the quark momentum. One then integrates over all 
such processes that can yield lepton pairs of beam-quark 
momentum fraction $x_1$ and invariant mass $M$. 
The inclusive cross section is given by
\begin{eqnarray}
{d\sigma_{DY}^{pN}(M^2)\over dx_1}=\int_{x_1}^1
dx_qF_q^p(x_q){d\sigma_{DY}^{qN}(M^2)\over dx_1^q}, \label{eq:signn}
\end{eqnarray}
where $F_q^p(x_q)$ is the quark distribution function of
the proton and $d\sigma_{DY}^{qN}(M^2)/dx_1^q$ is the quark-nucleon
differential cross section for lepton-pair production\cite{hir,bhq,kst1}.
A fit to the p-$^2H$ data yields the quark-nucleon cross section 
unmodified by energy loss or shadowing.

Moving to the description of the DY process on a nuclear target, we
consider two limits. In the first, characterized by modest values of 
$x_2$, the virtual photon is able to 
resolve individual nucleons of the nucleus. In the second, at very small 
values of $x_2$, in a manner 
analogous to shadowing in DIS described earlier, the virtual photon 
is able to resolve only clusters of nucleons, and in the extreme, only 
the full size of the nucleus. The transition 
between the two is controlled by the coherence length of the virtual 
photon\cite{kst1}, a measure of its resolving power. For the DY 
process it is given by
\begin{equation} 
l_c=\left\langle\frac{2\,E_q\,x^q_1\,(1-x^q_1)}
{(1-x^q_1)\,M^2+(x^q_1\,m_q)^2+k_T^2}\right\rangle\ , 
\label{eq:l_c} 
\end{equation}
where $E_q=x_q\,E_p$ and $m_q$ are the energy and mass of the projectile quark
which radiates the virtual photon.  The resulting lepton pair has an effective mass
$M$, a transverse momentum $k_T$, and carries a fraction $x^q_1$ of the
initial momentum of the quark.  The mean coherence length for
the kinematic conditions of E772 has been evaluated in Ref.~\cite{longpaper} 
by integrating over $x^q_1$ and $k_T$. Roughly speaking, energy loss is 
the dominant source of nuclear dependence when $l_c < 2$ fm, the average 
distance between nucleons in the nucleus. For $l_c > 2$ fm, shadowing 
predominates. 

Two features of the TRF formulation of the DY process, 
pioneered by Kopeliovich and collaborators\cite{hir,johnson,kp} are 
essential in the quantitative analysis of quark energy loss. First, 
shadowing may be calculated exactly (within the model assumptions) for 
both DIS and the DY process. The description for both is 
connected to the dipole cross section, $\sigma (\rho)$, for the 
absorption of a 
$q\bar q$ pair of transverse separation $\rho$. This phenomenology has 
been utilized extensively for high-energy photon reactions at 
HERA\cite{ac}. The second essential feature afforded by the TRF 
formulation is the determination of a more realistic path length for the 
projectile quark traversing the nuclear target. For example for tungsten
the mean path length is $\langle L\rangle =2.4$ 
fm, whereas for a uniform sphere $L_0=3R_0A^{1/3}/4 =4.9$ fm. This leads 
to larger values of $dE/dz$ derived from the data since there is a 
shorter path length in which to lose energy.

\section{Vacuum energy loss from DY nuclear dependence}

\begin{figure}                                              
  \begin{center}
    \mbox{\epsfxsize=2.25in\epsffile{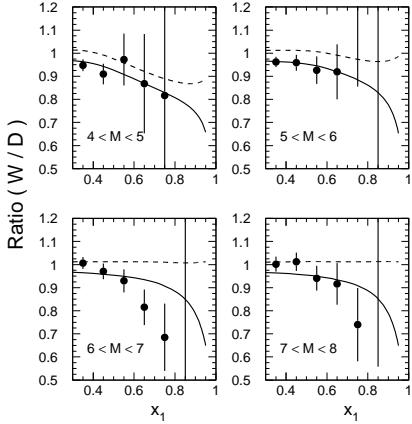}}         
    \vspace*{-9pt}                                     
  \end{center}
  \caption{Ratio of tungsten to deuterium
Drell-Yan yields per nucleon versus $x_1$
 for different intervals of $M$.
Dashed curves show the effect of shadowing. 
The solid curves include both 
shadowing and energy loss. Note that shadowing predominates for small 
masses, while the opposite obtains in the larger-mass bins.}
  \vspace*{-14pt}
  \label{w-d}
\end{figure}

With only one free parameter, the nuclear dependence 
ratios for $C$, $Ca$, $Fe$, and $W$ were fitted to yield $dE/dz$. 
A crucial feature of the fit is that it is performed on the 
nuclear-dependence ratios binned in both $x_1$ and $M$, since
energy loss and shadowing have contrasting kinematical features.
The fit yields a substantial energy loss, 
$-dE/dz = 2.32\pm 0.52\pm 0.5$ GeV/fm (statistical and systematic 
errors). Fits to $W/^2H$ in four mass intervals are shown by 
solid curves in Fig.~\ref{w-d}. 

\begin{figure}
\begin{center}     
    \mbox{\epsfxsize=3.0in\epsffile{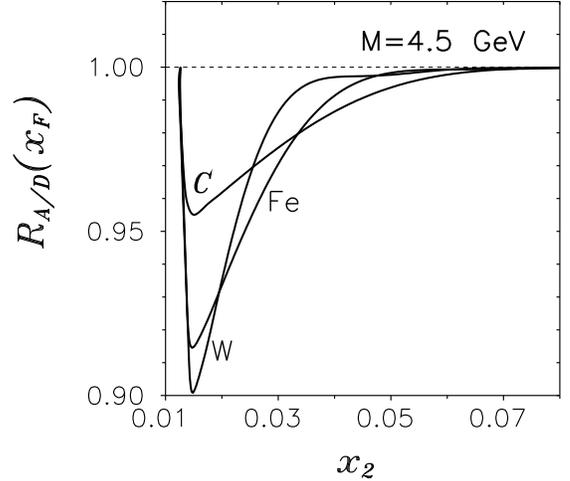}}             
\end{center} 
  \caption{Pure shadowing in the DY process in the present model as a function 
 of $x_2$ at M = 4.5 GeV.}
  \label{shadx2}
 \end{figure}

Our analysis, formulated in the TRF, differs significantly from 
previous energy-loss analyses\cite{e866,gm} (see 
Ref.~\cite{longpaper} for more detail). Among the most important 
differences is the treatment of shadowing. Fig.~\ref{shadx2} shows pure 
shadowing in the DY process for C, Fe, 
and W targets as a function of the target momentum fraction $x_2$. 
The phenomenology of Ref.~\cite{eks}, based on QCD evolution applied to 
DIS and DY shadowing data, and employed in the analysis of E866\cite{e866}, 
shows stronger shadowing for W than for C at all values of $x_2$. However 
the present analysis shows a very rapid decrease 
in shadowing for W at larger values of $x_2$. 
At $M$ = 6.5 GeV (corresponding to the lower-left frame in Fig.~\ref{w-d}) 
the DY process is limited by kinematics to $x_2\geq 0.028$. 
Here, shadowing is actually a larger effect for C than for 
W\footnote{To save space, we have not shown the corresponding 
calcuations at $M$ = 6.5 GeV; see Ref.~\cite{longpaper}}. 
Thus, in our analysis, the observed 
decrease in $R(W/D)$ seen at higher masses in Fig.~\ref{w-d} (lower two 
frames) is predominantly energy loss. 
Unfortunately current shadowing data in DIS cannot distinguish between 
these two competing phenomenologies, where the most dramatic differences 
are seen for very heavy targets.

The energy loss determined here should be interpreted as the {\it vacuum 
energy loss}. It has little to do with the medium itself, but is brought about 
by the first interaction which triggers hadronization. Induced energy 
loss is dicussed below.

\section{Induced energy loss}

Much theoretical attention has been devoted in recent years to the QCD 
analogue of the famous
LPM\cite{lp} effect\cite{bsz}.
It is now accepted that gluon 
radiation induced when a quark penetrates 
nuclear matter leads to additional energy loss 
proportional to the square of the path length traversed. This should lead to 
an observable broadening of the transverse momentum distribution given by,
\begin{eqnarray}
-dE/dz = {3\over 4}\alpha_s p_t^2.
\label{eq:lpm}
\end{eqnarray}
The measured $p_t$ broadening\cite{aldeups} of DY muon pairs from 
Tungsten is $\Delta p_t^2 = 0.1$ GeV$^2$ implying a maximum value 
$-(dE/dz)_{rad}\approx 0.2$ GeV/fm. However, this value should be 
considered approximate since the derivation of 
Eq.~\ref{eq:lpm} presumes the applicability 
of perturbative QCD, and the nuclear $p_t$ broadening effect is clearly
very small. Even so it is clear that induced energy loss in cold 
nuclear matter is not a large part of the total energy loss for 800 
GeV/c protons.

The present analysis has relied on the contrasting kinematical 
behavior of energy loss and shadowing to separate the two effects at 800 
GeV/c. It is clear in the present model that cold-matter energy loss, 
the effect of which scales as $\Delta E/E_p$ ($E_p$ being the laboratory 
beam energy in the TRF), will make a much 
smaller contribution in p-A collisions at RHIC energies. There, shadowing 
will be the dominant 
nuclear effect. On the other hand at lower beam energies, such as the 
120 GeV proton beam available at the Fermilab Main Injector (FMI), shadowing 
will be kinematically forbidden (for dilepton masses $\geq$ 4 GeV), with 
energy loss providing an even larger nuclear dependence to the DY 
process. Thus DY experiments at RHIC and the FMI are 
clearly very important in the ultimate clarification of these two 
important manifestations of QCD in bulk nuclear matter.




\end{document}